\documentclass[fleqn,usenatbib]{mnras}

\usepackage{newtxtext,newtxmath}

\usepackage[T1]{fontenc}
\usepackage{ae,aecompl}

\usepackage{graphicx}	
\usepackage{amsmath}	

\title[CALSAGOS]{CALSAGOS: Clustering ALgorithmS Applied to Galaxies in Overdense Systems}

\author[D. E. Olave-Rojas et al.]{
D. E. Olave-Rojas$^{1}$,\thanks{E-mail: daniela.olave@utalca.cl}
P. Cerulo$^{2}$,
P. Araya-Araya$^{3}$,
D. A. Olave-Rojas$^{4}$,
\\
$^1$Departamento de Tecnolog\'ias Industriales, Facultad de Ingenier\'{i}a, Universidad de Talca, Los Niches km 1, Curic\'o, Chile\\
$^{2}$Departamento de Ingenier\'{i}a Inform\'{a}tica y Ciencias de la Computaci\'{o}n, Universidad de Concepci\'{o}n, Chile\\
$^3$Departamento de Astronom\'{i}a, Instituto de Astronomia, Geof\'{\i}sica e Ci\^encias Atmosf\'ericas da Universidade de S\~ao Paulo,\\ Cidade Universit\'aria, CEP: 05508-900, S\~ao Paulo, SP, Brazil \\
$^{4}$Instituto de Investigaci\'on Interdisciplinaria, Universidad de Talca, 1 Poniente 1141, Talca, Chile\\
}

\date{Accepted XXX. Received YYY; in original form ZZZ}

\pubyear{2022}

\begin{document}
\label{firstpage}
\pagerange{\pageref{firstpage}--\pageref{lastpage}}
\maketitle
%
\begin{abstract}

In this paper we present \texttt{CALSAGOS}: Clustering ALgorithmS Applied to Galaxies in Overdense Systems which is a \texttt{PYTHON} package developed to select cluster members and to search, find, and identify substructures. \texttt{CALSAGOS} is based on clustering algorithms and was developed to be used in spectroscopic and photometric samples. To test the performance of \texttt{CALSAGOS} we use the S-PLUS's mock catalogues and we found an error of 1\% - 6\% on member selection depending on the function that is used. Besides, \texttt{CALSAGOS} has a $F_1$-score of 0.8, a precision of 85\% and a completeness of 100\% in the identification of substructures in the outer regions of galaxy clusters ($r > r_{200}$). The $F_1$-score, precision and completeness of \texttt{CALSAGOS} fall to 0.5, 75\% and 40\% when we consider all substructure identifications (inner and outer) due to the function that searches, finds, and identifies the substructures works in 2D and cannot resolve the substructures projected over others.
\end{abstract}




\section{Introduction}
In the last four decades, several authors (e.g. \citealt{1980Dressler}, \citealt{Couch1998}, \citealt{2004Kauffmann}, \citealt{Postman2005}, \citealt{Peng2010}, \citealt{2016Jaffe}, \citealt{2017Cerulo}, \citealt{2018OlaveRojas}, \citealt{2020vanderBurg}, \citealt{2021LimaDias}) have shown that the environment plays a fundamental role in the evolution of galaxies. In this context, theoretical and observational studies (e.g \citealt{2002lewis}, \citealt{poggianti06}, \citealt{2009mcgee}, \citealt{2010Demarco}, \citealt{2016Jaffe}, \citealt{2017Cerulo}) show that the some physical properties of galaxies are affected by the environment in which these are located and that the environment can trigger processes that drive the evolution of galaxies.

In particular, the observational evidence shows that galaxies in dense environments such as clusters and groups are less star-forming than galaxies in sparse environments (\citealt{poggianti99}, \citealt{dressler99}, \citealt{poggianti06}, \citealt{gobat08}). According to this, galaxy clusters are one the best laboratories to study and understand the environmental drivers of galaxy evolution. Due to the broad range of densities available in clusters, we can trace the evolution of galaxies through the analysis of galaxy properties from the diversely dense cluster outskirts to the dense cluster cores.

Besides, according to the $\Lambda$CDM hierarchical paradigm, galaxy clusters grow and increase their mass through the continuous accretion of less massive structures (i.e galaxies from the field or groups) or even through the merger with other clusters or filaments (\citealt{1974Press}, \citealt{2010Fakhouri}, \citealt{Chiang2013}). In this sense, the study of galaxy properties in the outer regions of galaxy clusters is crucial to understand the connections between galaxy evolution and the formation of their hosting large-scale structures. In particular, several astronomers (\citealt{2014Hou}, \citealt{2014Cybulski}, \citealt{haines2015}, \citealt{2018OlaveRojas}, \citealt{2019just}) have focused on the study of the fraction of star-forming galaxies in the outer regions of galaxy clusters. In general these works show that the fraction of quiescent galaxies (i.e galaxies without star-formation) is always higher in clusters than in the field even at large clustercentric distances ($r\gtrsim 3r_{200}$, e.g. \citealt{haines2015}, \citealt{2018OlaveRojas}).

The above results cannot be explained only by models in which quiescent galaxies are accreted directly from the field or by theoretical models in which galaxies quenched their star-formation when they crossed the cluster dark matter halo at $r_{200}$  \citep{haines2015}. For this reason, some authors (e.g. \citealt{2004fujita}, \citealt{2009mcgee}, \citealt{2013Vijayaraghavan}), suggest in their theoretical models that the pre-processing of galaxies \citep{1998Zabludoff}, is the responsible for the high fraction of quiescent galaxies in the outer regions of galaxy clusters. The pre-processing is a scenario in which galaxies quench their star-formation in galaxy groups before these systems are accreted onto a cluster (\citealt{2009mcgee}, \citealt{2013Vijayaraghavan}, \citealt{bahe13}). In this context, theoretical works (e.g. \citealt{2013Vijayaraghavan}) show that it is possible to detect galaxy groups that were accreted onto clusters as substructures within the three-dimensional distribution of the cluster member galaxies. As \cite{1988dressler} showed, galaxies in substructures retain part of the kinematic information about their progenitor groups.

Motivated by these findings, we have embarked in a project dedicated to the study of the effects of the environment on the evolution of galaxies, particularly focussing on the understanding of the connections between galaxy evolution and the formation of galaxy clusters. By dividing a cluster of galaxies into its main halo and its substructures, one can study the properties of galaxies in these two distinct environments and infer the relevance of pre-processing in shaping the physical properties of galaxies. 

The first difficulty to overcome in any kind of observational study of the pre-processing is the accurate detection of substructures and galaxy groups in and around galaxy clusters (e.g \citealt{Cohn2012}, \citealt{tempel17}). One observational technique for the identification of substructures is based on the analysis of the X-ray light profile of the Intra Cluster Medium (ICM) where substructures appear as peaks (e.g. \citealt{2018Bianconi}). However, this method is biased in favour of the most massive and dense substructures within a cluster.

In order to identify low- and high-mass substructures and accreted groups, some authors (e.g. \citealt{dressler13}, \citealt{2014Hou}, \citealt{2016Jaffe}, \citealt{2018OlaveRojas}, \citealt{2021LimaDias}) use the Dressler-Schectman Test (DS-Test, \citealt{1988dressler}). This statistical test allows one to determine the probability that a cluster has substructures and also assigns to each galaxy a value that corresponds to the probability of a galaxy to belong to a substructure. Galaxies with high probability to be a part of a substructure have peculiar velocities that deviate from the peculiar velocity distribution of the cluster. It is important to note that: 1) to effectively identify substructures with the DS-Test it is necessary to have a large spectroscopic coverage in the outskirts of the cluster, 2) the substructures identified are preferentially in the cluster outskirts (\citealt{1990West}, \citealt{2012Hou}, \citealt{dressler13}), and 3) the DS-Test only evaluates whether a galaxy belongs or not to a substructure with a certain probability without returning a list of the substructures of the cluster (\citealt{dressler13}, \citealt{2013Jaffe}). For the latter task it is necessary to combine the DS-Test output with other techniques based for example on the clustering of data (e.g \citealt{2018OlaveRojas}).

Clustering algorithms are unsupervised machine learning algorithms that were developed to group data points within a sample. So one can apply clustering in the three dimensional space of right ascension, declination and redshift, or in the two-dimensional space of right ascension and declination if spectroscopic redshifts are not available, to detect substructures in clusters of galaxies. One of the most known clustering algorithms is the mixture of Gaussians in which it is assumed that the distribution of galaxies is well reproduced by a collection of Gaussians, with each Gaussian corresponding to a single substructure (e.g. \citealt{2016Balestra}). Other clustering algorithms are based on the density of the data points in the parameter space. In these cases the clusters are defined as areas of high density separated by areas of low density. In this kind of algorithms the density of the clusters is generally defined based on the number of points and distance between them (e.g \citealt{2019Xu}). Due to their simplicity and versatility, in the last decade clustering algorithms have become popular as tools to identify substructures in clusters of galaxies (e.g. \citealt{2014Cybulski}, \citealt{2018OlaveRojas}, \citealt{2021LimaDias}) especially when there is low spectroscopic information.

This paper presents \texttt{CALSAGOS}: Clustering ALgorithmS Applied to Galaxies in Overdense Systems, which is a \texttt{PYTHON} package based on several clustering algorithms developed to select cluster members and to search, find, and identify substructures and galaxy groups in and around galaxy clusters. This package was developed to carry out the analysis presented by \cite{2018OlaveRojas}, and we are releasing it to provide the astronomy community with a tool that can be used for the detection of substructures in cases in which there is limited or no spectroscopic information available. Furthermore, this package can exploit both spectroscopic and photometric redshifts when they are available/measurable, with all the limitations carried with the usage of photometric redshift, as we will discuss in Section \ref{sec:substructure_identification}. Throughout the paper we adopt a CDM cosmology with $\Omega_\Lambda = 0.685$, $\Omega_m = 0.3$ and $h = H_0/100$ km s$^{-1}$ Mpc$^{-1}= 0.673$ \citep{2014Planck} in order to be consistent with the mock catalogues \citep{2021Araya_Araya} used to test \texttt{CALSAGOS}.

In this paper we will recurringly use the terms principal halo, subhalos, and substructures. The principal halo refers to the main body of the cluster. The subhalos correspond to structures smaller than the principal halo that are linked to it; the members of each subhalo are gravitationally linked between them and bounded to its potential well. The substructures are galaxy groups that were accreted into the principal halo and that can be distinguished within the main body of the cluster through the analysis of the velocity and spatial distribution of the galaxies.

The paper is organized as follows. In Section \ref{sec:structure} we present and describe \texttt{CALSAGOS}. In Section \ref{sec:test} we present the testing and results of the performance of \texttt{CALSAGOS}. In Section \ref{sec:discussion} we discuss the performance and limitations of \texttt{CALSAGOS}. Finally, in Section \ref{sec:summary} we summarise our main conclusions.

\section{Structure of CALSAGOS}\label{sec:structure}

\texttt{CALSAGOS} was developed in \texttt{PYTHON} version 3.8, using pre-existing modules such as \texttt{numpy} version \texttt{`1.21.3'}, \texttt{astropy} version \texttt{`4.3.1'}, \texttt{matplotlib} version \texttt{`3.4.3'}, \texttt{sys}, \texttt{math}, \texttt{sklearn} version \texttt{`1.0.1'}, \texttt{scipy} version \texttt{`1.7.1'} and \texttt{kneebow} version \texttt{`0.1.1'}. 

\subsection{Modules}

\texttt{CALSAGOS} has seven modules that can be used separately or together, depending on user preferences. In each module there are several functions written to estimate different galaxy properties and select cluster members, and/or assign galaxies to substructures. The modules of \texttt{CALSAGOS} are listed below:
\begin{enumerate}
    \item \texttt{utils}: This module contains a collection of utility functions developed to estimate errors, physical quantities of the clusters, projected distances, and convert units.
    \item \texttt{redshift\_boundaries} This module has a collection of functions developed to estimate the central cluster redshift, the velocity dispersion of the cluster, and limits of the redshift distribution, which can be used to select the cluster members.
    \item \texttt{cluster\_kinematics} This module contains functions that allow the user to estimate some kinematic properties of a single galaxy cluster.
    \item \texttt{ds\_test} contains a collection of functions developed to implement the DS-Test. This module is recommended for use when there are spectroscopic data.
    \item \texttt{ISOMER} Identifier of SpectrOscopic MembERs is a module that contains a function to identify spectroscopic cluster members. These are defined as those galaxies with a peculiar velocity lower than the escape velocity of the cluster. This module was developed to carry out the analysis presented in \cite{2018OlaveRojas}.
    \item \texttt{CLUMBERI} CLUster MemBER Identifier is a module that contains a function that identifies cluster members assuming that the sample is well reproduced by a collection of Gaussian. \texttt{CLUMBERI} can select spectroscopic or photometric cluster members through a 3D implementation of the Gaussian-Mixture Models (GMM, \citealt{2010muratov}) and using the positions and redshifts of the galaxies.
    \item \texttt{LAGASU} LAbeller of GAlaxies within SUbstructures is a module that allows one to assign galaxies to different substructures in and around galaxy clusters by using GMM and Density-Based Spatial Clustering of Applications with Noise (DBSCAN, \citealt{ester}).
\end{enumerate}

For more details about the different modules and functions please read the help of each one. The webpage of \texttt{CALSAGOS}\footnote{\url{https://github.com/dolaver/calsagos/}} also provides an example that shows the functioning of the software.

\subsection{Clustering Algorithms}

The identification of substructures in clusters of galaxies is performed in two steps which are the determination of cluster membership and the identification of substructures in the two- or three-dimensional spatial distribution of cluster member galaxies. We describe the two steps in this sub-section, considering the case in which spectroscopic or photometric redshifts are available and the case in which no redshift information is available. It is important to stress that if we are provided with a catalogue of cluster members it is not necessary to run again the first step.

\subsubsection{Members Selection}

To select cluster members using the spectroscopic redshifts of the galaxies we have developed \texttt{ISOMER}, which is a function that selects cluster members as those objects with a peculiar velocity lower than the escape velocity of the cluster. The escape velocity of the cluster is estimated using the equation published in \cite{1999diaferio} given by:
\begin{equation}\label{eq1}
    v_{esc} \simeq 927 \left(\frac{M_{200}}{10^{14} h^{-1} M_\odot} \right)^{1/2} \left(\frac{r_{200}}{h^{-1} \text{Mpc}} \right)^{-1/2} \text{km s}^{-1}
\end{equation}

To use this equation we need to know the $M_{200}$ and $r_{200}$\footnote{$M_{200}$ and $r_{200}$ are respectively the mass and radius of a sphere with a mean density equal to 200 times the critical density of the Universe at the redshift of the cluster.} of the cluster. If one only knows $M_{200}$, $r_{200}$ can be derived using the equation published in \cite{2005finn}:
\begin{equation}\label{eq2}
    200 \rho_c (z) = \frac{M_{cl}}{4/3 \pi r^3_{200}}
\end{equation}

Where $\rho_c(z)$ corresponds to the critical density of the Universe at a particular redshift. Finally, the peculiar velocity of each galaxy is estimated with the following equation published in \cite{1974harrison}:
\begin{equation}\label{eq3}
    v = c \frac{z- z_{cl}}{1+ z_{cl}}
\end{equation}

This equation is valid, to first order, for $v << c$. Equations (\ref{eq1}), (\ref{eq2}) and (\ref{eq3}) are all implemented in \texttt{cluster\_kinematics}. 

In the practice, to select cluster members all galaxies with a peculiar velocity higher than the escape velocity of the cluster are removed from the sample. Then the velocity dispersion is estimated in this clean sample using the biweight estimator described in \cite{1990Beers} and that is implemented in  \texttt{astropy}. The error in the velocity dispersion is determined through bootstrap resampling. Finally, the field interlopers are removed by using a 3$\sigma$ clipping algorithm \citep{1977Yahil}. This approach was developed and implemented in \cite{2016Cerulo} and \cite{2018OlaveRojas}. The uncertainties in the cluster member selection will be discussed in detail in the next section.

When the spectroscopic information is only partially available or not available at all, clustering methods can be adopted to select cluster members and substructures. \texttt{CLUMBERI} is a function that uses the GMM implementation of \texttt{sklearn} to select cluster members when the spectroscopic and/or photometric redshifts and the positions of galaxies in the sky are available.

GMM is a clustering algorithm that is based on the assumption that the distribution of data points in the parameter space of a sample is well represented by a collection of $n$ Gaussian functions, each function representing a cluster. The points in the sample are assigned to a certain Gaussian according to their probability of belonging to it. \texttt{CLUMBERI} requires as input the maximum number \textit{N} of Gaussians to fit for four different types of co-variances, i. e.\ \textit{spherical}, \textit{tied}, \textit{diag}, \textit{full}. Thus, for each definition of the covariance matrix, \texttt{CLUMBERI} runs GMM with a number of Gaussians that grows from 1 to \textit{N} and then uses the Bayesian Information Criterion (BIC, \citealt{1978Schwarz}) to select the best model. It is important to note that \textit{n} corresponds to the number of Gaussians that reproduce the sample and \textit{N} corresponds to the maximum number of Gaussians that we can input to GMM. So, when GMM finds the best model, it selects a number of \textit{n} Gaussians that is between 1 and \textit{N}. Finally, the fit of the Gaussians to the galaxy distribution is performed with the Expectation-Maximization (EM, \citealt{1977dempster}, \citealt{2007press}) algorithm.

\texttt{CLUMBERI} selects cluster members using a 3D implementation of GMM in the space defined by the equatorial coordinates (i.e.\ R.A. and Dec.) and the redshift of the galaxies. Starting from the initial values of the central position and redshift of the cluster (e.g.\ those that are reported in a catalogue), \texttt{CLUMBERI} compares the position  of the maximum of each Gaussian with the initial position of the centre of the cluster and selects the Gaussian closest (in redshift and position) to the centre of the cluster. Once \texttt{CLUMBERI} has identified the galaxies in this Gaussian, the position of the centre and the redshift of the cluster are updated, and these updated values are used to select cluster members. More specifically, all the galaxies that are at more than 3$\sigma$ from the updated redshift of the cluster centre are rejected. After the rejection of the outliers \texttt{CLUMBERI} updates the redshift of the cluster and searches for outliers at more than 3$\sigma$ from the updated value of the cluster redshift. The procedure stops when there are no galaxies at more than 3$\sigma$ from the cluster redshift. Finally, \texttt{CLUMBERI} estimates the number of cluster members, errors in the estimate of membership, redshift, the velocity dispersion of the cluster, and the uncertainty in velocity dispersion. The errors in the velocity dispersion of the cluster are estimated with bootstrap. The uncertainty in the number of cluster members will be discussed in detail in the next section.

\subsubsection{Substructure Identification}\label{sec:substructure_identification}

When one has a reliable estimation of cluster membership, it is possible to proceed to grouping galaxies within substructures in and around clusters. For this purpose, we developed \texttt{LAGASU}, which is based on two unsupervised Machine Learning algorithms, namely GMM and DBSCAN, both available in \texttt{sklearn},

DBSCAN is a clustering algorithm in which a group is defined as a collection of points within a high-density area separated by low-density areas. To assign the points of the sample to different groups, DBSCAN requires as input the position of the points, epsilon ({\ttfamily{eps}} or $\epsilon$), and the {\ttfamily{min\_samples}}. The {\ttfamily{eps}} parameter corresponds to the distance threshold for points to be considered as neighbours: all points at distances shorter than {\ttfamily{eps}} from a point are its neighbours. {\ttfamily{min\_samples}} corresponds to the minimum number of neighbours that a point must have within {\ttfamily{eps}} to be considered as member of a real group. All points outside the identified groups are considered as noise (for further details please see Section 3 of \citealt{ester})

The \texttt{LAGASU} module has two functions, namely \texttt{lagasu} and \texttt{lagasu\_dbscan}. The \texttt{lagasu} function assigns galaxies to substructures using GMM and DBSCAN. \texttt{lagasu} first uses GMM to model the redshift distribution of galaxies with a mixture of $n$ Gaussians. The best number of Gaussians is determined as in \texttt{CLUMBERI}, using the BIC. Afterwards the redshift range of cluster members is sliced according to the width of the Gaussians into $n$ bins. \texttt{lagasu} automatically runs DBSCAN in each redshift bin to detect substructures in the projected distribution of the galaxies and assigns galaxies to the substructures encountered. The \texttt{lagasu\_dbscan} function assigns galaxies to different substructures using only DBSCAN.
   
If the user has access to spectroscopic data we suggest using \texttt{lagasu} because DBSCAN only works in the two-dimensional space of galaxy coordinates. We recommend this since there exist cases in which two galaxies are close in projection but far in redshift. However, if the user has only access to the photometric data we recommend \texttt{lagasu\_dbscan} because the errors in photometric redshift may be high. For instance, in the case of S-PLUS \cite{2020Molino} found that the estimated $z_{phot}$ have a precision of 3\% for galaxies with $z <$ 0.5, while \cite{2021Balogh} reports an accuracy of 5\% in the Gemini Observations of Galaxies in Rich Early ENvironments (GOGREEN) survey at $0.8 \leq z < 1.5$. As a comparison, the precision of the spectroscopic redshifts in GOGREEN varies between $5 \cdot 10^{-4}$ and $2 \cdot 10^{-3}$. 

It is important to note that we do not use GMM to identify the substructures because this algorithm assigns all galaxies to one substructure/group. So there could exist a case in which two galaxies very distant spatially are assigned to the same group. DBSCAN, on the other hand, does not assign all galaxies in the sample to one group \citep{ester}, and galaxies that are not assigned to a substructure are classified as ``noise'' and assigned to the principal halo of the cluster.

We remark that GMM and DBSCAN are two clustering algorithms, and therefore these tools cannot be used to automatically distinguish between the principal halo and the substructures. All structures are classified as groups. For this reason we suggest the following criterion: when having a spectroscopic sample, the group nearest to the cluster centre should be considered as the principal halo. When only photometric data are available, all groups within a projected distance of $r_{200}$ from the cluster centre should be considered as part of the principal halo. This implementation is done in the \texttt{test.py} published on the web-page of \texttt{CALSAGOS}.

Finally, if the user has access to the spectroscopic data, they  can first run \texttt{ds\_test} over the sample of cluster galaxies and then \texttt{lagasu}. This with the aim of finding substructures that have kinematic deviations with respect to the cluster.

\section{Testing the performance of CALSAGOS}\label{sec:test}

\subsection{Data: Mock galaxy catalogue}\label{sec:test_data}
To test the performance and quantify the accuracy of \texttt{CALSAGOS}, we use the same mock catalogue created for \cite{2022bWerner}, who similarly probed the galaxy cluster detection in the Data Release 1 (DR1) of the Southern Photometric Local Universe Survey (S-PLUS, \citealt{2019MendesdeOliveira}). The lightcone, constructed with the techniques in \citet{2021Araya_Araya}, has a projected area of 324 deg$^2$ and goes up to $z \sim 0.5$. Here we briefly describe its construction for completeness.

The synthetic galaxies of this mock come from applying the L-GALAXIES semi-analytical model (SAM,  \citealt{2015Henriques}) to the Millennium Run (MR) simulation \citep{2005Springel} scaled to the \textit{Planck 1} cosmology \citep{2014Planck} (using the \citealt{2010Angulo} algorithm, included in the SAM source code). The cosmological redshifts are attributed to each galaxy assuming that sources at comoving distances $d_c(z_i) \leq d_{c,gal} \leq d_c(z_i) + 30 $ kpc are at the same redshift $z_i$. Afterwards, the ``observed" redshifts are computed by considering the peculiar motions of the galaxies along the line of sight. Finally, following \cite{2007Kitzbichler}, the celestial coordinates, namely right ascension and declination, are obtained.

The spectro-photometric properties of the modelled galaxies were obtained following the \textit{post-processing} approach described in \cite{2015Shamshiri}. This consists in using the star formation history (SFH) arrays of each galaxy (L-GALAXIES output) whose bins store information about stellar masses and metallicities produced between two cosmic times for the following baryonic components: disk, bulge, and intra-cluster medium (ICM). SFH arrays have as many as 20 times (age) bins, depending on the simulation snapshot. For a description of the binning algorithm, see \cite{2015Shamshiri}. The sum of disk and bulge SFH arrays can represent the whole stellar component of the modelled galaxies, where each bin can be associated with a simple stellar population and have a spectral energy distribution (SED). The used SED templates are from the \cite{2005Maraston} stellar synthesis population models, assuming a \cite{2003Chabrier} initial mass function (IMF). Over the galaxy component SEDs (sum of all SFH bin SEDs) it is applied the dust extinction model following the approach used by \cite{2007DeLucia}, \cite{2015Henriques}, \cite{2015Shamshiri}, \cite{2015Clay}. Then, the final galaxy SED is the sum of both bulge and disk components after the dust correction. Apparent magnitudes were computed for the five S-PLUS broad-bands ($u$, $g$, $r$, $i$, and $z$ filters), by placing each galaxy SED in the observer-frame, according to its ``observed'' redshift.

\subsubsection{Simulated clusters and their substructures}

 We did not know a priori which galaxies are cluster/group members. To identify them, we correlated the mock data with a catalogue with dark matter halo information. The latter was obtained from the Virgo-Millennium \footnote{\url{http://gavo.mpa-garching.mpg.de/MyMillennium/}} database. Here, as well as in \cite{2022bWerner}, we defined the simulated clusters as all primary dark matter halos with $M_{200} \geq 10^{14} \ M_{\odot}$. In practice (following the database notation), from the \texttt{MPAHaloTrees..MRscPlanck1} table, we selected all halos with \texttt{haloId} $=$ \texttt{firstHaloInFOFgroupId} and $\texttt{m\_crit200} \geq 10^{14} \ M_{\odot}$. Notice that, at this point, the first condition would select just galaxies in the main cluster halo. We estimated the central Right Ascension (R.A.), Declination (Dec.), and redshift of each cluster as the median position of the galaxy members in its main halo. 

At this point, to identify the galaxies in subhalos linked to a particular cluster with \texttt{haloId} $=$ $i$ in the mock, again from the Virgo-Millennium database, we extracted all the dark matter halos with \texttt{firstHaloInFOFgroupId} $= i$. By running this search iteratively for all galaxy clusters in the mock, we finally obtained a catalogue with all bounded subhalos and thus galaxies in substructures.

In our implementation of \texttt{CALSAGOS}, we defined a substructure/group as a collection of at least three neighbouring galaxies. For this reason, we removed all those subhalos in the mock with two or fewer members. To test the performance of \texttt{CALSAGOS} we used 200 halos in the mass range $14.7 \leq \log(m_{200}/M_\odot) \leq 15.5$ and in the redshift range $0.038 \leq z \leq 0.575$.


\subsection{Initial Configuration}\label{sec:initial_config}

\begin{figure}
\includegraphics[width=\columnwidth]{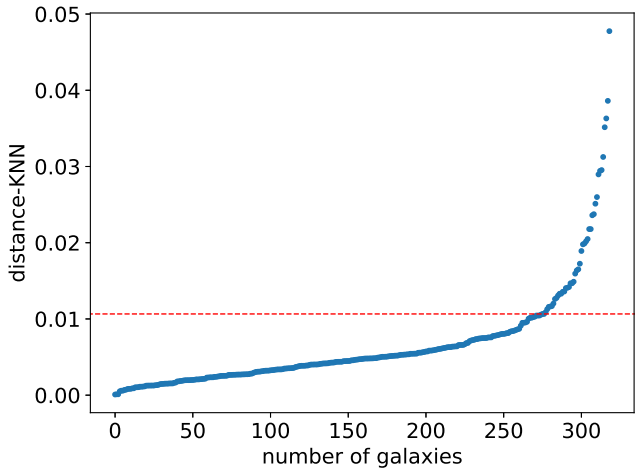}
\caption{Sorted distance-KNN graph for the sample of the halo ID26 in our clean S-PLUS mock catalogue using K = 3. The horizontal red dashed line corresponds to the threshold value of the distance between galaxies. All galaxies with a higher distance-KNN value are considered to be noisy and all other galaxies are assigned to some substructure.}
\label{dbscanplot}
\end{figure}

In the test.py available in the \texttt{CALSAGOS} repository, a group/substructure is defined as a collection of at least three neighbouring galaxies. This is implemented by setting the parameter {\ttfamily{n\_galaxies}} (or {\ttfamily{min\_samples}}) to 3. However, the user can set this parameter as they prefer as long as its value is greater than 1. The {\ttfamily{eps}} parameter was determined following the approach suggested by \cite{ester} and \cite{Rahmah2016}. In practice, we estimated the {\ttfamily{eps}} in each cluster by measuring the distance of each galaxy to its K-Nearest Neighbour (KNN), in our case $K = 3$ since we set {\ttfamily{n\_galaxies}} = 3. Then we sorted the distances in ascending order and selected the {\ttfamily{eps}} as the point of maximum curvature which was determined through an automated procedure that uses the functions \texttt{calc\_knn\_galaxy\_distance} and \texttt{best\_eps\_dbscan} from the module \texttt{utilis} of \texttt{CALSAGOS}. In Figure \ref{dbscanplot} we can see the plot of the sorted KNN-distance for a selected random halo from the S-PLUS mock catalogue (identified as ID26 in our clean catalogue) using K = 3. As it is mentioned above, we selected the {\ttfamily{eps}} as the distance in which the plot has its maximum curvature. This value is represented by a red dashed line in the plot. All galaxies with a KNN-distance above the red line are considered to be noise and all galaxies below this value are assigned to some substructure.

We set to 15 the maximum number of Gaussians that can be fitted in \texttt{CLUMBERI} and \texttt{LAGASU}. We tested the performance of \texttt{CALSAGOS} setting the maximum number \textit{N} of Gaussians to be fitted in \texttt{CLUMBERI} and \texttt{LAGASU} to values varying from 5 to 30 in steps of 5, and we found similar values of the BIC regardless of the number of Gaussians used. The user can set the parameter \textit{N} to the same number in both \texttt{CLUMBERI} and \texttt{LAGASU} because in both functions the value of $n$ Gaussians is determined with the BIC in an independent way.

We have tested the performance of \texttt{lagasu} and \texttt{lagasu\_dbscan} and found the same results for both approaches. For this reason, hereafter \texttt{LAGASU} corresponds to the \texttt{lagasu} function.

\subsection{Uncertainties in the performance of \texttt{CALSAGOS}}\label{sec:errors}

To estimate the error on our cluster member selection we ran \texttt{CALSAGOS} in the sample defined in Section \ref{sec:test_data} and we estimated the relative difference between the members of the principal halo in the mock and the members selected by using \texttt{CLUMBERI} or \texttt{ISOMER}. Then we estimated the median of the distribution of the relative differences. 

To quantify the performance of \texttt{LAGASU} in the selection of substructures we estimated the $F_1$-score \citep{1979VanRijsbergen} which allows one to quantify the accuracy in the analysis of binary classification. Our method for the detection of substructures determines whether a galaxy belongs or not to a substructure, so it can be interpreted as a binary classification problem in which we are classifying galaxies into main cluster members and substructure members. \texttt{CALSAGOS} addresses this problem with unsupervised learning, which does not allow one to determine the numbers of false negatives and false positives. However, we chose to test our software on the S-PLUS mock catalogue in which the positions, redshifts and structures of the dark matter halos are all known. This sample provides us with a ground truth against which we can test the performance of the clustering procedures implemented in \texttt{CALSAGOS}; therefore, the $F_1$-score can be used in this case to quantify the accuracy of our method for selecting substructures in clusters of galaxies.

The $F_1$-score quantifies the quality of the performance of \texttt{LAGASU} as the number of true positive identifications divided by the sum of the number of true positive identifications and the number of total false identifications (i.e. the false negative plus the false positives). This corresponds to the harmonic mean of precision and recall. The $F_1$-score is defined as:

\begin{equation}
    F_1 = \frac{t_p}{t_p + \frac{1}{2}(f_p + f_n)}
\end{equation}

where $t_p$ corresponds to the \textit{true positive} identifications, $f_p$ is the \textit{false positive} identifications and $f_n$ corresponds to the \textit{false negative} identifications. The true positives are substructures identified by \texttt{LAGASU} that correspond to subhalos in the mock catalogue. False positive identifications are substructures that were identified by \texttt{LAGASU} but do not correspond to subhalos in the mock catalogue. Finally, false negative identifications correspond to subhalos that were not identified as substructures by \texttt{LAGASU}. The values of the $F_1$-score range from 0 to 1, where 1 corresponds to a perfect classification and 0 means that there is total disagreement between the true and the assigned classes. In addition, the precision or purity of \texttt{LAGASU} was defined as:
\begin{equation}
    P = \left(\frac{t_p}{t_p + f_p}\right)100\%,
\end{equation}
where $P$ ranges from 0 to 100\% and the completeness or recall was defined as:
\begin{equation}
    C = \left(\frac{t_p}{t_p + f_n}\right)100\%.
\end{equation}
$C$ also ranges between 0 and 100\%.

\subsection{Results}\label{sec:results}

Using the initial configuration described in Section \ref{sec:initial_config} we estimate an error of 1\% and 6\% in the member selection when we use \texttt{CLUMBERI} and \texttt{ISOMER}, respectively. This difference in the errors is due to \texttt{ISOMER} always selecting fewer cluster members than \texttt{CLUMBERI}. Despite the difference in the number of members obtained with these two functions, when we applied \texttt{LAGASU} over the sample of cluster members, the output was the same regardless of using \texttt{CLUMBERI} or \texttt{ISOMER}.

When we consider all substructures (inner and outer) we obtain a $F_1$-score of 0.5, a precision of 75\% and a completeness of 40\%. When we analyse the type of identifications we find that 64\% of subhalos are false negatives. Therefore, we are not able to identify the 64\% of the subhalos. This is due to the fact that there is information in 3D that we have not considered. However, 73\% of our identifications are true positives (i.e. subhalos that are identified as substructures). These results are mainly due to the fact that \texttt{LAGASU} cannot identify substructures projected over others or over the main cluster, because this module is based on clustering algorithms that work in the two-dimensional coordinate space.

For the reason mentioned above we analyse the performance of \texttt{LAGASU} only considering the outer substructures (i.e substructures outside $1 \times r_{200}$ from the cluster centre), and we obtain a F$_1$ score of 0.8, a precision of 85\% and a completeness of 100\%. When we analyse the type of identifications, we find that 85\% of them are true positives. Furthermore, when considering only the outer substructures, none of the subhalos are false negatives. In other words, all subhalos in the outer regions of the cluster are identified as substructures. On the other hand, when we analyse the performance of \texttt{LAGASU} only considering the inner substructures (i.e substructures within $1 \times r_{200}$ from the cluster centre) we obtain a $F_1$-score of 0.4, a precision of 80\% and a completeness of 30\%. In this case, the 80\% of our identifications are true positives and the 70\% of the subhalos are false negatives.

The $F_1$ and the percentages of true positive and false negative were obtained by estimating each value for each cluster and then deriving the median of the distribution of each of these values for the whole sample. To estimate the error in the number of members of the substructures we used the same approach used to estimate the error in the cluster members, and we found 30\% and 20\% when considering all and only the outer substructures, respectively.

\begin{figure*}
\includegraphics[width=\columnwidth]{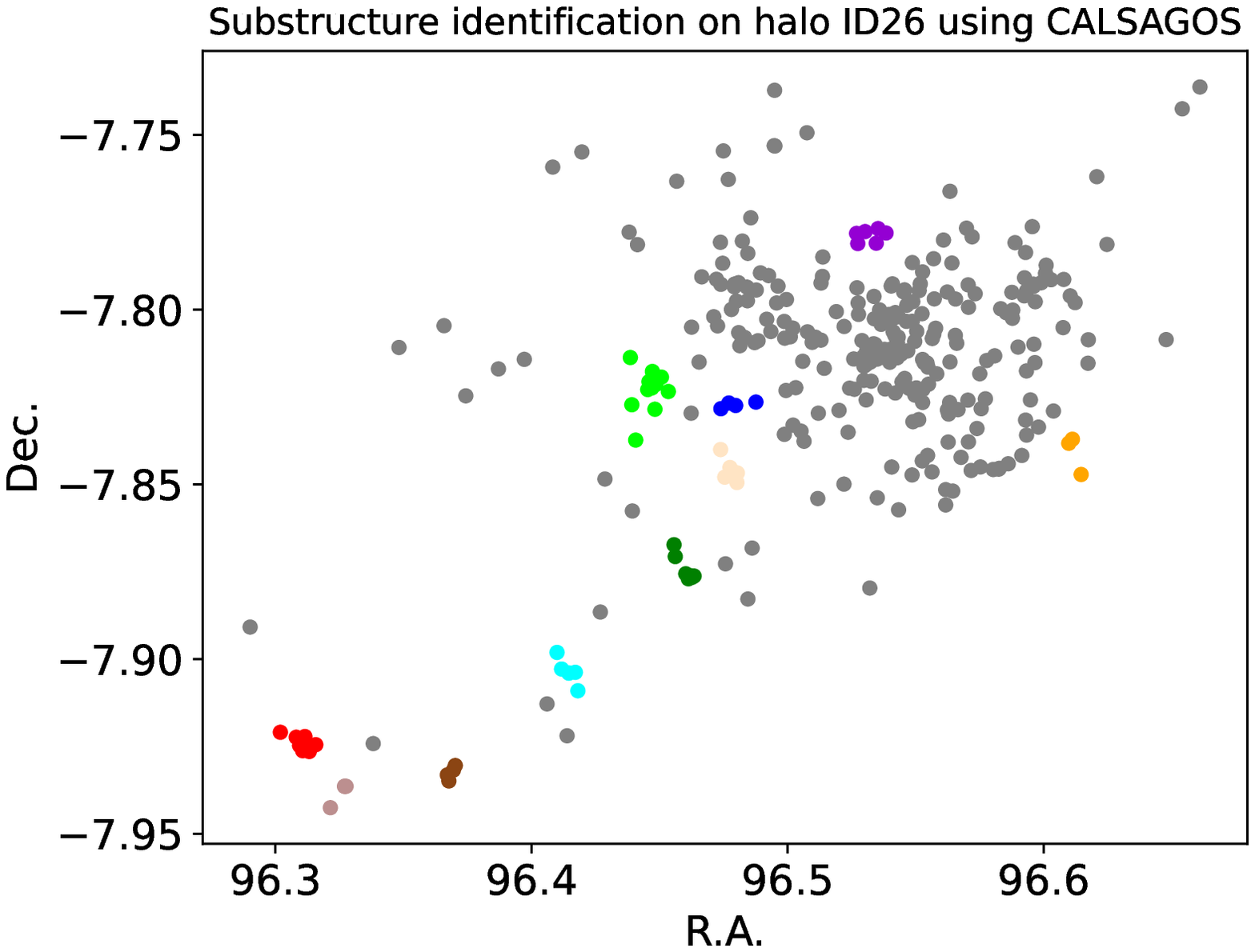}
\includegraphics[width=\columnwidth]{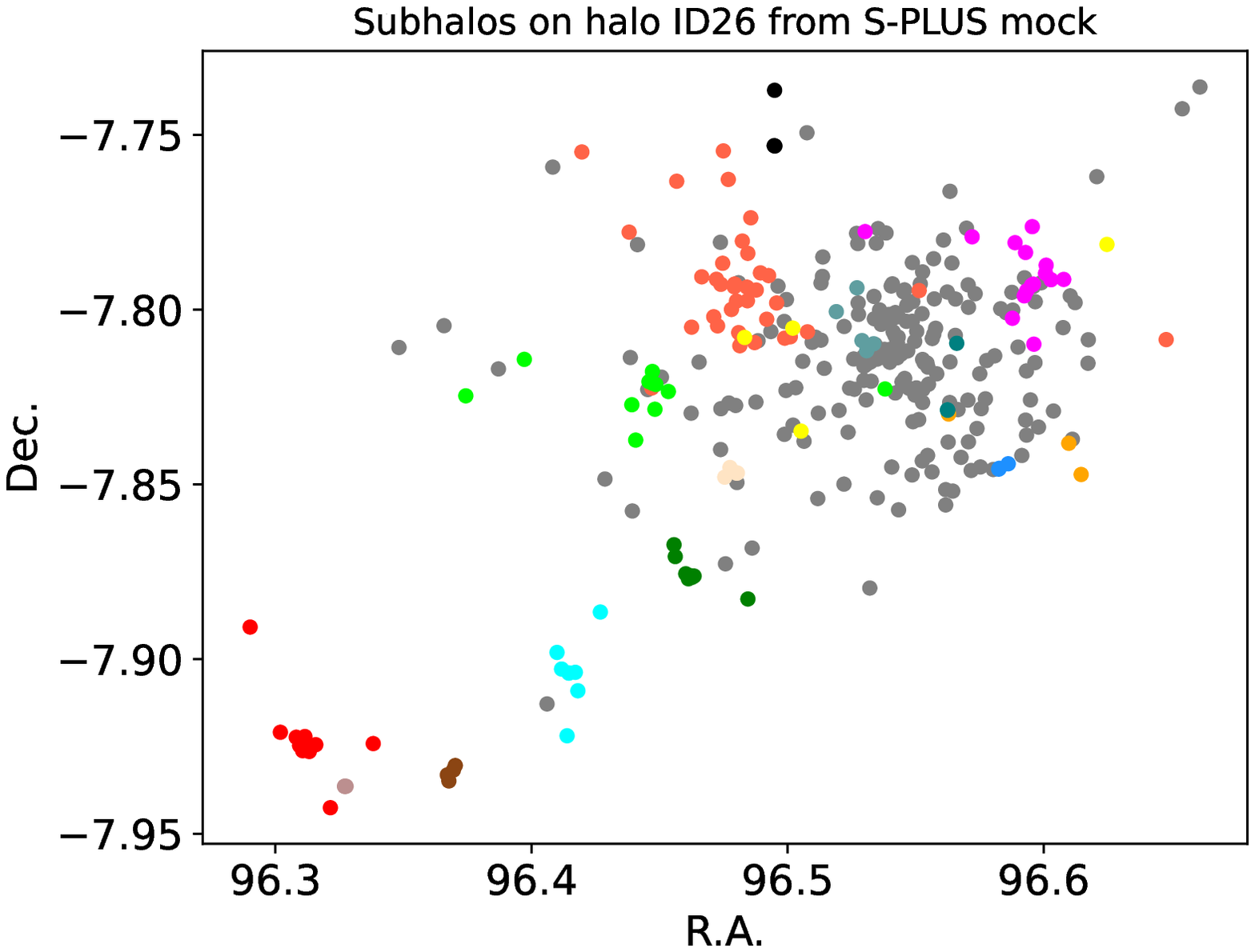}
\caption{Left panel: substructure identification after running CALSAGOS in the halo ID26 from the S-PLUS mock. Right panel: subhalos in the ID26 halo. According to the S-PLUS mock, the selected halo has 16 subhalos with 3 or more members. With \texttt{CALSAGOS} we identified 11 substructures in this halo. In both cases galaxies in the principal halo are plotted with grey dots and galaxies in subhalos/substructures are plotted with dots of different colours. Galaxies with the same colours in both panels correspond to the same substructure/subhalos.}
\label{output_calsagos}
\end{figure*}

In Figure \ref{output_calsagos} we present the output of \texttt{CALSAGOS} for the halo ID26. In the left panel we show the cluster member selection and substructure identification. In this example cluster members were selected with \texttt{CLUMBERI}. According to the S-PLUS mock catalogue, this halo has 319 members and 16 subhalos with 3 or more members. \texttt{CALSAGOS} detects 319 $\pm$ 3 members and 11 substructures, which were defined using the criteria for the spectroscopic sample. In the right panel of the Figure \ref{output_calsagos} we show the halo and subhalos according to the information taken from the mock catalogue. The true positive substructures in the left panel have the same colour of the corresponding subhalo in the right panel. Instead, false positive substructures and false negative subhalos are coloured with different colours. Finally, galaxies that are part of the cluster (or principal halo) but are not part of a substructure (or subhalo) are coloured grey.

\begin{figure*}
\includegraphics[width=\textwidth]{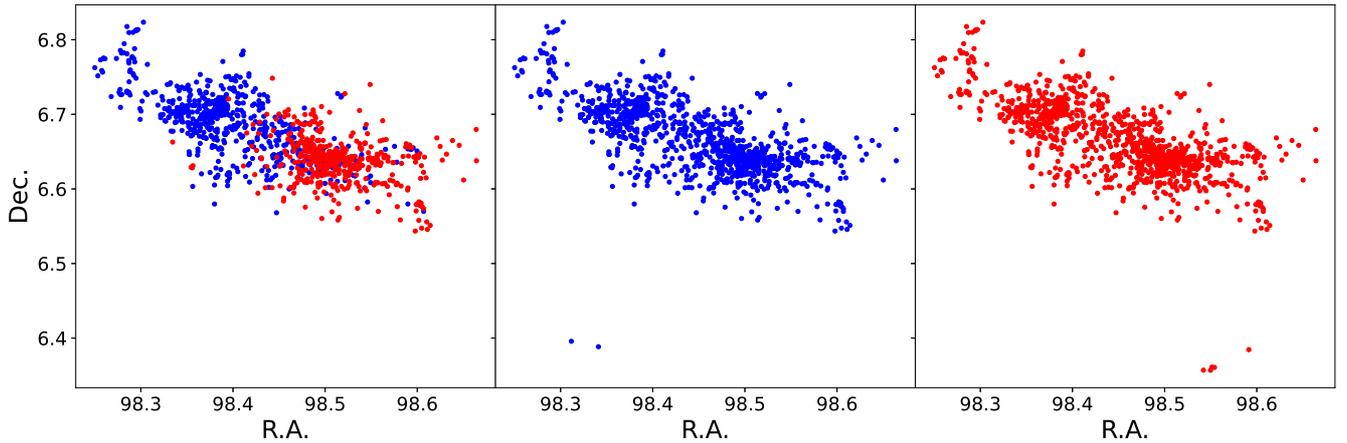}
\caption{Left panel: blue points represent the members of halo ID4 taken from the mock. This halo has 682 members. Red points represent the members of halo ID5 taken from the mock. This halo has 503 members. Middle panel: all blue points correspond to the members of halo ID4 selected using \texttt{CLUMBERI}. Right panel: all red points correspond to the members of halo ID5 selected using \texttt{CLUMBERI}. We obtained 1185 and 1190 members for the halos ID4 and ID5, respectively.}
\label{output_2_calsagos}
\end{figure*}

The S-PLUS mock catalogue contains the halos ID4 and ID5 that are very close to each other in position and redshift (see Table \ref{table_output_calsagos}). When we plot the members of these halos (see left panel of Figure \ref{output_2_calsagos}) we can observe that one halo overlaps with the other as if the halos were merged. Due to this fact, regardless of whether we use \texttt{CLUMBERI} or \texttt{ISOMER} to select cluster members of one of these halos, we always select both halos as if they were one (see middle and right panel of Figure \ref{output_2_calsagos} and Table \ref{table_output_calsagos}). Therefore, we find an error of 70\% when we selected members using \texttt{ISOMER} and an error of 130\% when we select members using \texttt{CLUMBERI}. Furthermore, when we run \texttt{LAGASU} over our catalogue of members obtained with \texttt{CLUMBERI} for these halos, we obtain a $F_1$-score of 0.2 and 0.1 for halo ID4 and ID5, respectively, a precision of 55\% and 20\% for halo ID4 and ID5, respectively and a completeness of 10\% for both halos. These results are considering all substructures (inner and outer).

\begin{table*}
\centering
\begin{minipage}[t]{\textwidth}
	\caption{Member identification halos ID4 and ID5}
 		\begin{tabular}{lcccccccccc}
\hline
		Halo    & R.A. & Dec. & $z$     &  mass    &  members\footnote{number of members according to the mock catalogue} & subhalos\footnote{number of subhalos with three or more members taken from the mock catalogue} & members\footnote{number of members obtained using \texttt{CLUMBERI}} & substructures\footnote{number of all substructures (inner and outer) obtained using \texttt{LAGASU} after selecting cluster members with \texttt{CLUMBERI} } & members\footnote{number of members obtained using \texttt{ISORMER}} & substructures\footnote{number of all substructures (inner and outer) obtained using \texttt{LAGASU} after selecting cluster members with \texttt{ISOMER} }   \\
		        & \multicolumn{2}{c}{(J2000.)}  &   & $\log(m_{200}/M_\odot)$         &  & & & & &    \\ 
		\hline
		ID4  &  98:26:22.04 &  06:39:25.73 & 0.454 &  15.21  & 682 & 51 & 1185 $\pm$ 12 & 11 $\pm$ 2 & 1129 $\pm$ 11 & 23 $\pm$ 5   \\
		ID5  &  98:30:06.98 &  06:38:24.94 & 0.455 &  15.19  & 503 & 33 & 1190 $\pm$ 12 & 21 $\pm$ 4 & 1133 $\pm$ 11 & 47 $\pm$ 9 \\
\hline
\vspace{-0.8cm}
\label{table_output_calsagos}
\end{tabular}
\end{minipage}
\end{table*}

\begin{figure*}
\includegraphics[width=\textwidth]{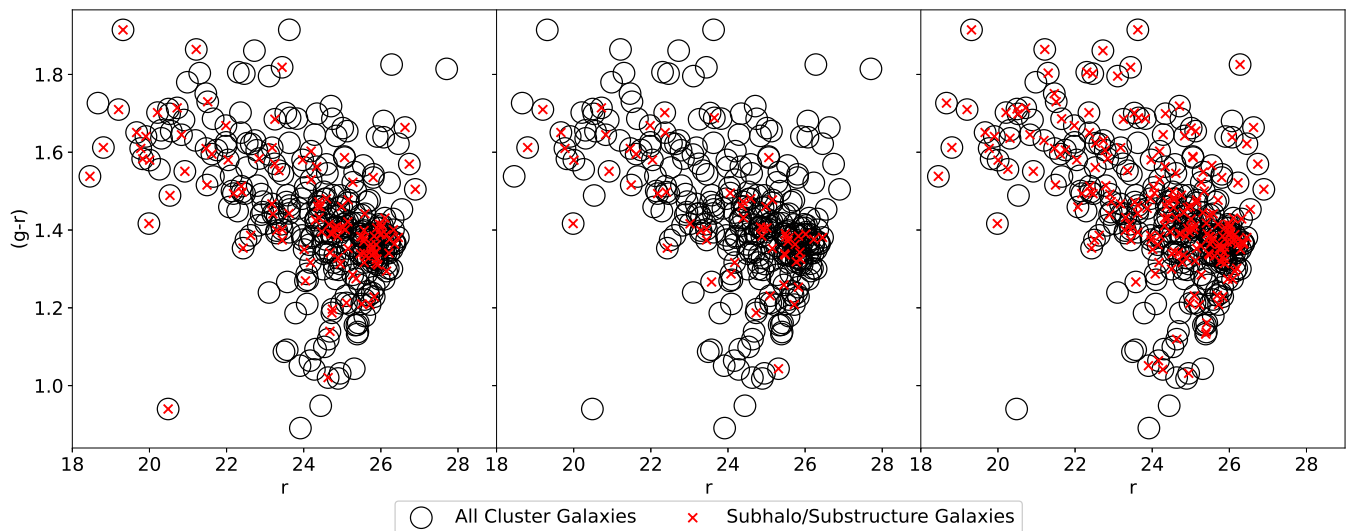}
\caption{Colour–magnitude diagrams of the halo ID26. The colours are measured in the \textit{g} and \textit{r} S-PLUS broad bands (see Section \ref{sec:test_data}). Cluster member galaxies are plotted with open black circles and galaxies in substructures are plotted with red crosses. Left panel: cluster members and subhalos of the halo ID26 taken from the mock. Middle panel: cluster members of halo ID26 was selected using \texttt{CLUMBERI} and substructure identification was done using \texttt{LAGASU}. Right panel: members of halo ID26 selected using \texttt{ISOMER} and substructure identification was done using \texttt{LAGASU}.}
\label{output_3_calsagos}
\end{figure*}

\begin{figure*}
\includegraphics[width=\textwidth]{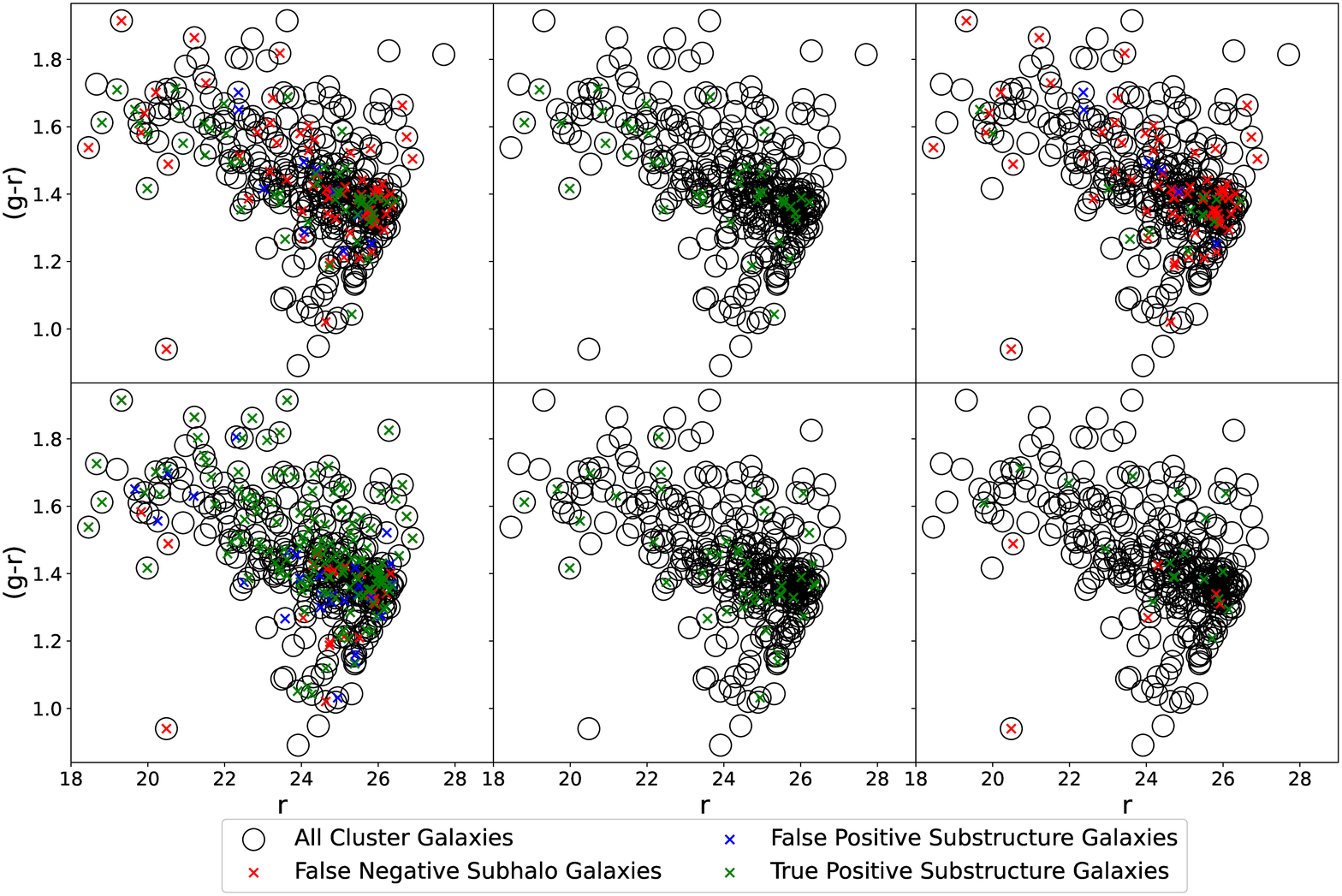}
\caption{Colour–magnitude diagrams of the halo ID26. The colours are measured in the \textit{g} and \textit{r} S-PLUS broad bands (see Section \ref{sec:test_data}). Cluster member galaxies are plotted with open black circles. Galaxies in false negative subhalos are plotted with red crosses, galaxies in false positive substructures are plotted with blue crosses and galaxies in true positive substructures are plotted with green crosses. In all panels the substructures where identified using \texttt{LAGASU}. In all top panels, the clusters members of halo ID26 which selected using \texttt{CLUMBERI} and in all bottom panels, the clusters members of halo ID26 which selected using \texttt{ISOMER}. Left panels: Identification of all substructures (inner and outer). Middle panels: identification of outer substructures (i.e. $r > r_{200}$). Right panels: identification of inner substructures (i.e. $r < r_{200}$)}
\label{output_4_calsagos}
\end{figure*}

However, if we compared our member selection with the combined halos (i.e. ID4 + ID5) as if they were a single halo, we obtain an error of 0\% in the member selection with \texttt{CLUMBERI} and \texttt{ISOMER}. Additionally, when we compare the results obtained with \texttt{CLUMBERI} and \texttt{LAGASU} for the halos ID4 and ID5 with the combined halos we obtain a $F_1$-score of 0.1, a precision of 55\% and a completeness of 7\% for the halo ID4, while for halo ID5 we obtain a $F_1$-score of 0.3, a precision of 70\% and a completeness of 20\%.

When we run only \texttt{LAGASU} over the catalogue with the members of halos ID4 and ID5 taken from the mock we obtain for halo ID4 a $F_1$-score of 0.4, a precision of 70\% and a completeness of 25\%, while for ID5 a $F_1$-score of 0.3, a precision of 70\% and a completeness of 20\%. Finally, we run \texttt{LAGASU} on a catalogue with the members of the combined halos and obtain a $F_1$-score of 0.6, a precision of 70\% and a completeness of 50\%.

\begin{table*}
\centering
	\caption{Performance of \texttt{CALSAGOS} on halo ID26 as a function of redshift type}
 		\begin{tabular}{l|ccc|ccc|ccc}
\hline
		    & \multicolumn{3}{c}{z$_{spec}$}  & \multicolumn{3}{c}{z$_{phot}$} & \multicolumn{3}{c}{without redshift}  \\
		    & all & outer & inner & all & outer & inner & all & outer & inner \\
		\hline
		F$_1$-score  &  0.67 & 1 & 0.53 & 0.86 & 0.8 & 0.64 & 0.67 & 1 & 0.53\\
		Precision &  82\% & 100\% & 80\% & 79\% & 75\% & 58\% & 82\% & 100\% & 80\%\\
		Completeness &  56\% & 100\% & 40\% & 94\% & 86\% & 70\% & 56\% & 100\% & 40\%\\
\hline
\vspace{-0.8cm}
\label{table_performance_calsagos}
\end{tabular}
\end{table*}

In order to discern if there is a bias due to colour or magnitude in the selection of cluster members and the detection of substructures with \texttt{CALSAGOS}, we used the apparent magnitudes in the mock catalogue (see Section \ref{sec:test_data}) to obtain the (\textit{g} - \textit{r}) colours of the mock galaxies and investigate any difference between the colour distributions of the subhalo galaxies and the galaxies in the main halo and between the substructure galaxies and the galaxies in the main cluster. Figure \ref{output_3_calsagos} shows the  (\textit{g} - \textit{r}) versus \textit{r} colour–magnitude diagram (CMD) for the halo ID26. In the left panel of Figure \ref{output_3_calsagos} we present the cluster members and subhalos taken from the mock and in the middle and right panels we show the cluster members and substructure identification using \texttt{CLUMBERI}+\texttt{LAGASU} and \texttt{ISOMER}+\texttt{LAGASU}, respectively.

Furthermore, in Figure \ref{output_4_calsagos} we present the CMD for the halo ID26, highlighting the true positive (green crosses), false positive (blue crosses), and false negative (red crosses) in our substructure identification. We also present the CMD for all substructures (left panels), outer substructures (middle panels), and inner substructures (right panels). In the top panels cluster members are selected using \texttt{CLUMBERI}, while in the bottom panels they are selected with \texttt{ISOMER}.

With the aim to discern if actual substructure galaxies follow the colour distribution of the cluster members, we perform the Kolmogorov-Smirnov Test (KS Test, \citealt{1933Kolmogorov}, \citealt{1948Smirnov}) to compare the colour distribution between the cluster members and substructure members. We use the implementation of the test contained in \texttt{ks\_2samp} \citep{Hodges1958} from the \texttt{PYTHON} module \texttt{scipy}. We compare the colour of subhalo galaxies with the colours of cluster members in the mock, and we find a p-value of 0.75. When we compare the colours of the galaxies in substructures with the colours of the cluster members selected with \texttt{CLUMBERI} and \texttt{ISOMER} we find a p-value of 1 and 0.2, respectively.

When we run the KS test to compare the colour distributions of all substructures with our member selection we find a p-value of 1, 0.6 and 0.3 for the false negative, false positive and true positive identifications, respectively. When we compare the colour distribution of the galaxies in the outer substructures with that of galaxies in the main cluster we find a p-value of 0.7. Finally, when we perform the comparison for the galaxies in the inner substructures we find a p-value of 0.6, 0.6 and 0.07, for the false negative, false positive and true positive, respectively. In all cases, the p-value is greater than 0.05 for a 95\% confidence level. We can conclude that there is no difference and therefore no induced bias in the colour distribution of galaxies in substructures.

Finally, to analyse the performance of \texttt{CALSAGOS} as a function of the different redshift definitions available, we performed the following tests on the halo ID26: i) we selected cluster members, and we identified substructures using the spectroscopic redshifts; ii) we selected cluster members, and we identified substructures using the photometric redshifts; iii) we identified substructures using only the projected positions (R.A. and Dec.). We used \texttt{LePHARE} (\citealt{1999Arnouts}, \citealt{2006Ilbert}) to estimate the photometric resdhifts in halo ID26, and the results of this test are summarised in Table \ref{table_performance_calsagos}.

Cluster members were selected using \texttt{CLUMBERI}, obtaining 319 $\pm$ 3 and 400 $\pm$ 4 cluster members using $z_{spec}$ and $z_{phot}$, respectively. Table \ref{table_performance_calsagos} shows that the precision of \texttt{CALSAGOS} in identifying substructures falls by 3\% for all substructures, 25\% for outer substructures and 22\% for inner substructures when we use $z_{phot}$. We expected the worst precision when we only used R.A. and Dec.; however, it can be seen that we obtain the same precision as when using $z_{spec}$. We argue that this could be a consequence of the fact that to perform this experiment we selected the halo and not a light cone. So, in this case the mock galaxies seen in projection coincide with the actual members of halo ID26.

\subsubsection{\texttt{CALSAGOS} on real data}\label{sec:results_reals}

The first version of \texttt{CALSAGOS} was probed in two real clusters at $z \sim 0.4$, namely MACS0416 and MACS1206, as part of the analysis published in \citealt{2018OlaveRojas}. In this case, we used: 1) \texttt{ISOMER} to select cluster members, 2) the DS-Test to verify the existence of substructures in the clusters and to verify the kinematic deviation of each galaxy from the main cluster, and 3) \texttt{LAGASU} to assign galaxies to different substructures (see Figure 2 in \citealt{2018OlaveRojas}). We found more substructures in the outer part ($r \geq r_{200}$) of the clusters than in the inner regions ($r < r_{200}$). These results agree with the results reported by other authors for these clusters (see \citealt{2013Biviano}, \citealt{2016Balestra}). We note that with \texttt{CALSAGOS}, we are able to recover the ``Sext'' structure identified by \citep{2016Balestra} in MACS0416.

In particular, for MACS0416 \citealt{2016Balestra} found 781 spectroscopic members and a velocity dispersion of $\sigma_{cl} = 996^{+12}_{-16}$ km s$^{-1}$, and these authors report a spectroscopic completeness that varies between 0.5 and 0.7 decreasing from the inner to the outer regions of the cluster. If we select the cluster members using \texttt{ISOMER}, we obtain 890 $\pm$ 53 spectroscopic members and a velocity dispersion of $\sigma_{cl} = 1101 \pm 22$ km s$^{-1}$. Our estimation of the velocity dispersion is consistent within $1\sigma$ with the result reported by \citet{2016Balestra}. Furthermore, when we select the cluster members using \texttt{CLUMBERI} we obtain 1036 $\pm$ 10 spectroscopic members and a velocity dispersion of $\sigma_{cl} = 1036 \pm 10$ km s$^{-1}$. The velocity dispersion estimated with \texttt{CLUMBERI} differs by $9\sigma$ from the value reported in \citet{2016Balestra}. However, we note that when we use \texttt{CLUMBERI} we find 32\% more spectroscopic members than \citet{2016Balestra} and the velocity dispersion estimated with \texttt{CLUMBERI} is consistent within 2$\sigma$ with the value estimated when we select the cluster members using \texttt{ISOMER}. Finally, it is important to note that the galaxies in the substructures of MACS0416 follow the same colur-magnitude relation as the whole cluster as was shown in Figure 3 of \citet{2018OlaveRojas}.

\begin{figure}
\includegraphics[width=\columnwidth]{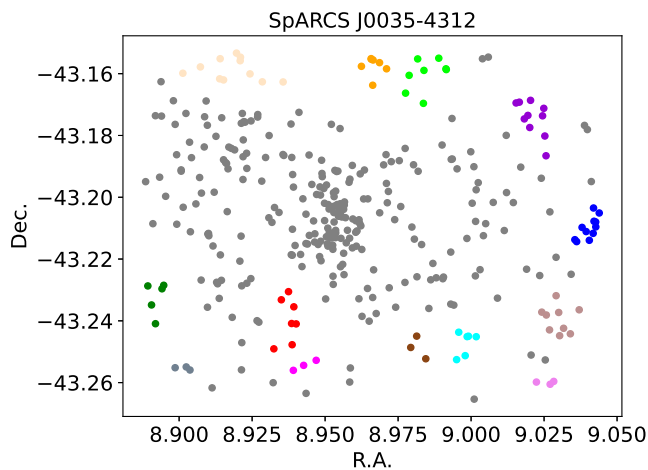}
\caption{Substructure identification in the SpARCS J0035-4312 cluster at $z=1.34$. Grey points correspond to galaxies in the principal halo of the cluster (i.e. galaxies that are cluster members but are not part of a substructure). Coloured points correspond to galaxies in different substructures.}
\label{gogreen}
\end{figure}

We also ran \texttt{CALSAGOS} in the cluster SpARCS J0035-4312 at $z=1.34$ drawn from the GOGREEN sample. The results are shown in Figure \ref{gogreen} in which points with different colours correspond to galaxies in different substructures and grey points correspond to cluster members that are not part of a substructure. For this case, we have used the criterion that all groups/substructures within $r_{200}$ from the cluster centre are considered as part of the principal halo.

\section{Discussion and Conclusions}\label{sec:discussion}

\subsection{The importance of the clustering algorithms on the substructure analysis}

Pre-processing has been demonstrated to be relevant in interpreting the results of galaxy evolution studies at low and high redshift (e.g. \citealt{2012Li}, \citealt{haines2015}, \citealt{2018OlaveRojas}, \citealt{2021Reeves}, \citealt{2022aWerner}). Thus, the study of galaxy properties in substructures and infalling groups in and around galaxy clusters is fundamental to constrain the importance of pre-processing in the quenching of star formation and structural transformations as well as investigate the links between galaxy evolution and the formation of large-scale structures. However, to carry out all these studies it is necessary to precisely identify the substructures and infalling groups in clusters of galaxies, especially at large cluster-centric distances ($r>r_{200}$).

There is no unique method to observationally search, find and identify substructures. For example, some authors use the projected phase-space diagram (e.g. \citealt{1982Geller}, \citealt{2013Biviano}), the X-ray emission of the ICM (e.g. \citealt{1992Briel}, \citealt{2018Bianconi}), the kinematics of galaxies (e.g. \citealt{dressler13}, \citealt{2014Hou}), the density of the galaxies (e.g. \citealt{2014Cybulski}) or combined methods (e.g. \citealt{2018OlaveRojas}, \citealt{2021LimaDias}). However, none of these methods can be considered the best and, in general, the performance of each method depends mainly on the kind of data, their quality, and the dynamical state of the clusters. In all the cases, the performance of a certain method improves when the statistics are high and when one considers parameter spaces with high dimensions (e.g. the performance improves when one adds redshifts to the positions of the galaxies) (\citealt{2007Ramella}, \citealt{tempel17}). Furthermore, as shown in \cite{Cohn2012}, substructures are less likely to be found in relaxed clusters than in merging clusters.

The majority of studies about pre-processing have so far been focused on the analysis of spectroscopic samples. This is mainly due to the fact that the methods used to search for substructures involve the derivation of peculiar velocities from the redshifts of the cluster members. The uncertainties of photometric redshifts are too high to derive peculiar velocities, and so methods such as the DS-test cannot be applied in these cases. However, this only allows the analysis of little samples of clusters (except in the case of the Local Cluster Substructure Survey, \citealt{haines2015}) because spectroscopic observations are time-consuming, and the collection of a large spectroscopic sample demands large observing times, especially at high redshifts (e.g. \citealt{2021Balogh}).

On the other hand, it is less time-consuming to conduct deep imaging observations in different filters that allow large multi-band photometric samples to be built. One can thus estimate photometric redshifts by fitting an SED to the fluxes at different wavelengths, increasing the statistics in the redshift of the galaxies. However, photometric redshifts are less precise than spectroscopic redshifts, and this does not permit the application of the method based on the estimation of peculiar velocities in the detection of substructures. For his reason, the photometric surveys of galaxy clusters have been little explored in the study of pre-processing. Clustering represents, therefore, a technique that can be used as a tool to reliably search for substructures when there is little or no spectroscopic information available. 

Clustering has been extensively used in astronomy to study the bimodality in the colour distribution of galaxies, to select cluster members or even to identify substructures. Among the most used algorithms ther are the Kaye’s Mixture Model (KMM, \citealt{1994Ashman}) and the Gaussian Mixture Model (GMM, \citealt{2010muratov}). Both algorithms group galaxies assuming that their distribution in position, redshift or colour is better represented by a collection of $n$ Gaussians rather than a single Gaussian. In this kind of algorithms the objects are assigned to groups according to their likelihood of membership. When we used Gaussian mixture methods to select substructures in and around galaxy clusters we noted that all the galaxies were assigned to one group that corresponded to the main cluster. This means that the algorithm would attribute galaxies that are not part of the same structure for their peculiar velocities and positions to the same group (e.g. \citealt{2013Jaffe}, \citealt{2018OlaveRojas}).

When we use photometric redshifts, this can be an even bigger problem due to the errors on the redshift estimation. In this case it is better to use clustering algorithms based on density of points to identify substructures. For this reason, it is necessary to use combined clustering algorithms or at least to select the best clustering algorithm to our purpose and test its performance in a sample with a known substructure list. 

\subsection{Strengths and limitations of CALSAGOS}

We observe that the selection with \texttt{ISOMER} has greater uncertainty than the selection with \texttt{CLUMBERI}. This could be due to the fact that \texttt{ISOMER} was developed to select spectroscopic cluster members. Indeed, on one hand we assumed a certain shape of the gravitational profile to use the escape velocity for the selection of cluster members, and on the other hand, when we use spectroscopic redshifts we always select fewer cluster members because sampling is always partial (e.g. see the spectroscopic observational strategies in \citealt{2013Biviano} and \citealt{2016Balestra}). In this sense, we can say that \texttt{ISOMER} selects cluster members more conservatively than \texttt{CLUMBERI}. However, \texttt{ISOMER} gives us a selection with less contamination than \texttt{CLUMBERI}.

Furthermore, when we use \texttt{LAGASU} to identify substructures in clusters we note that DBSCAN searches for groups without distinguishing between the principal cluster and the substructures. This means that when we run \texttt{LAGASU} over the galaxy cluster, the central part of the cluster (i.e. the principal halo in the mock) is also selected as a group within the sample. This is because DBSCAN does not receive in input any information about the dynamics of the galaxy cluster. For this reason, we have decided to add two criteria: 1) when we use a spectroscopic sample, the group that is nearest to the central position and redshift of the cluster corresponds to the main cluster, and the rest of the identified groups are considered as substructures, and 2) when we use photometric samples, all groups in the central part of the cluster (i.e. within $r_{200}$) are considered as part of the principal halo.

To quantify the precision of \texttt{CALSAGOS} in the detection and mapping of galaxy cluster substructures in observational data, we ran the package in a region of the mock catalogue without considering the information about the membership of the galaxies to the subhalos. When we consider all substructures (inner and outer) we obtain a $F_1$-score of 0.5 and, precision of 75\% and a completeness of 40\%. The low value of the $F_1$-score is not necessarily due to a failure in the package, but rather to its limitations, because \texttt{LAGASU} assigns galaxies to each substructure by using the two-dimensional position of the galaxies (R.A. and Dec.) and the number density.  This produces a high number of false negatives, i.e. the program does not find a high number of substructures that are in fact subhalos in the mock. Interestingly, the majority of these false negatives are in the inner part of the halo where projection effects may affect more the detection of substructure and so \texttt{LAGASU} does not resolve them.

When we only consider the outer substructures (i.e. $r > r_{200}$) we obtain a $F_1$-score of 0.8, a precision of 85\% and a completeness of 100\%. In particular, 85\% of the identifications are true positive and 0\% of the subhalos are false negative. When we only consider inner substructures (i.e. $r \leq r_{200}$) the $F_1$-score drops to 0.4. This value is a result of the fact that 70\% of the identified subhalos are false negative. Nevertheless, 80\% of our identifications are true positives. Therefore, the low performance in the inner part of the cluster is a result of the fact that in this region the substructures overlap with the main cluster, generating a projection effect and the consequent confusion between the main cluster and the substructures. This effects is less important in the cluster outskirts and, as a result, the performance of \texttt{LAGASU} increases when we only consider the outer substructures. Besides, due to physical mechanisms within the cluster core, internal groups and substructures are more likely to be destroyed and are thus more difficult to detect (\citealt{2013Vijayaraghavan}; \citealt{2013Jaffe}).

The above results are obtained independently of the method used for the selection of cluster members (\texttt{ISOMER} or \texttt{CLUMBERI}), and we also obtain the same results when we select a halo from the mock and only run \texttt{LAGASU} over it. Besides, it is important to note that the methods used with observational data to search, find and identify substructures differ from those used in numerical simulations. In the case of the observational data, the substructure is defined in terms of position and kinematic. In the case of numerical simulations, the galaxies that belong to a subhalo must be gravitationally bound to one another. Thus, if \texttt{CALSAGOS} identifies a substructure in a cataglogue from a cosmological simulation in which the subhalos are known, this may not always correspond to a subhalo.

According to the results of the $F_1$-score, it can be concluded that the method presented in this paper has a residual intrinsic bias that has to be taken into account whenever {\ttfamily{CALSAGOS}} is used to select cluster members and substructures. In particular, we cannot obtain most of the information on the substructures that are closer to the cluster centre in observational data. However, the projected information in two dimensions tells us about the distribution of galaxies in the cluster, and we note that a substructure in the cluster could or not fall into a selected group by the clustering algorithms.

The results obtained from the detailed analysis of halos ID4 and ID5 show that \texttt{CALSAGOS} is not able to distinguish clusters that are in the process of merging. However, if we treated these systems as one we can reach a $F_1$-score of 0.6, a precision of 70\% and a completeness of 50\%. The $F_1$-score is above the median value obtained for the entire sample considering all the substructures (see Section \ref{sec:results}), the precision is below the median and the completeness is above the median.

Figure \ref{output_3_calsagos} and the results of the KS test demonstrate that our method for cluster member selection and substructure identification do not introduce biases in colour and apparent magnitude. Thus, the mock galaxies have statistically indistinguishable colour distributions regardless of being cluster, subhalo or substructure members. On the other hand, Figure \ref{output_4_calsagos}, together with the results of the KS test, shows that mock galaxies in all, outer or inner identified substructures have the same ($g-r$) colour distribution of the mock galaxies in the entire cluster. Furthermore, we note that the galaxies in the false negative subhalos, false positive and true positive substructures follow the same distribution as the cluster members in the CMD. This result is also corroborated by the results of the KS test. According to this, we can conclude that \texttt{CALSAGOS} does not introduce colour or magnitude dependent biases both in the selection of cluster members and in the identification of substructures.

On the other hand, Table \ref{table_performance_calsagos} shows that the precision of \texttt{CALSAGOS} in the substructures identification depends on the quality of the cluster member selection and the type of redshift: if we use photometric redshifts to select cluster members, the precision is lower than when using $z_{spec}$. Onthe other hand, the completeness tends to be better when we use $z_{phot}$ because more galaxies are selected as cluster members.

In spite of the fact that we used the S-PLUS mock catalogues to test the performance and accuracy of \texttt{CALSAGOS}, we want to remark that this package was also successfully employed in observational data of clusters of galaxies at intermediate and high redshift to search, find and identify substructures within and around galaxy clusters as we show in Section \ref{sec:results_reals}. Thus this paper demonstrates that \texttt{CALSAGOS} can be used on real or mock data to reliable detect substructures in galaxy clusters and conduct studies on the pre-processing of galaxies.

\section{Summary}\label{sec:summary}

So far the majority of the observational works about the pre-processing and the properties of galaxies in the outer regions ($r \geq r_{200}$) of clusters have been developed on sparsely populated samples because they were focused on spectroscopic data. This is mainly due to the techniques used in the selection of cluster members and in the tools used to search for substructures. However, clustering algorithms are powerful tools that can be employed in the analysis of substructures in clusters of galaxies when only photometric redshifts or no redshifts are available. As shown in \cite{2018OlaveRojas}, this allows one to increase the statistics in these kinds of studies ad to overcome the limitations of the spectroscopic samples.

For these reasons, the development of \texttt{CALSAGOS} offers an opportunity to extend the analysis of galaxy properties to the outer regions of clusters of galaxies by using photometric samples, because this is a tool based on clustering algorithms that has been tested on a mock galaxy sample which eliminates the problem of the dependence on spectroscopic data.

We find that the DBSCAN algorithm used in \texttt{LAGASU} provides a reliable detection of the cluster substructures. In particular we find that \texttt{LAGASU} has a better performance when identifying substructures in the outer regions of the cluster ($r \geq r_{200}$) reaching a $F_1$-score of 0.8,  a precision of 85\% and a completeness of 100\%. When we consider all the substructures the $F_1$-score, the precision and the completeness drop to 0.5, 70\% and 40\%, respectively. We attribute this  to the fact that in the inner regions of the clusters it is more likely to have several substructures along the same line of sight. As a result, \texttt{LAGASU}, which works in two dimensions, cannot resolve them because they appear as the same substructure in projection. This is the main limitation of \texttt{CALSAGOS}. 

We must keep in mind that we probe the performance of an observational method to search, find and identify substructures over a mock catalogue in which we have subhalos that were defined according to their gravitational potential. However, when one only works with photometric samples in which no spectroscopic redshift is available, the information on the dynamics of the galaxies in a cluster cannot be obtained. So \texttt{LAGASU} works without prior knowledge of the dynamics of the cluster, and this means that the substructures that are identified may not correspond to systems with internal dynamics that differ from those of the principal cluster.

\section*{Acknowledgements}
We thank the anonymous referee for the helpful and constructive feedback. We especially thank Claudia Mendes de Oliveira the P.I. of the Southern Photometric Local Universe Survey (S-PLUS) for giving us access to the mock catalogues to test this package. We are also grateful to Ricardo Demarco, Yara Jaff\'e, and Diego Pallero for their input on this project. Besides, we would like to thank everyone who has tested \texttt{CALSAGOS} extensively and provided us with invaluable feedback. PA-A thanks Coordenação de Aperfeiçoamento de Pessoal de Nível Superior (CAPES) for supporting his PhD scholarship (project 88887.596140/2020-00)

\section*{Data Availability}
\texttt{CALSAGOS} can be downloaded from the webpage \url{https://github.com/dolaver/calsagos} that also contains the working tutorial \texttt{test.py}. Besides, the mock catalogue used in this work is available upon request from Pablo Araya-Araya (e-mail: paraya-araya@usp.br).

\bibliographystyle{mnras}
\bibliography{main.bib} 

\bsp	
\label{lastpage}
\end{document}